%% file: 25ESSERC_MAED.tex
\documentclass[10pt,conference,twocolumn]{IEEEtran}

\usepackage{cite}
\usepackage[T1]{fontenc}
\usepackage{graphicx}
\usepackage{amssymb}
\usepackage{amsmath}
\usepackage{amsthm}
 
\usepackage{microtype}
\usepackage{balance}
\usepackage{subfigure}
\usepackage{xcolor}
\usepackage{booktabs}
\usepackage{soul}
\usepackage{algorithm}
\usepackage{tikz}

\usepackage{algorithmicx}
\usepackage{algpseudocode}
\algrenewcommand\algorithmicindent{0.7em}%
\usepackage{xpatch}
\usepackage{dblfloatfix}
\usepackage{framed}
\usepackage{setspace}

\input{macros/vmr-symbols-vecbold}
\input{macros/standard-macros}

\input{macros/defs}

\definecolor{myred}{HTML}{9E292B}
\definecolor{myblue}{HTML}{235787}
\definecolor{mygreen}{HTML}{5E6638}
\definecolor{mygray}{HTML}{444444}
\definecolor{myblack}{HTML}{000000}
\definecolor{mywhite}{HTML}{FFFFFF}

\definecolor{myaltred}{HTML}{D46A78}
\definecolor{myaltblue}{HTML}{6699C2}
\definecolor{myaltgreen}{HTML}{B0B58C}
\definecolor{myaltgray}{HTML}{AAAAAA}

\definecolor{mylightred1}{HTML}{B15455}
\definecolor{mylightred2}{HTML}{C57F80}
\definecolor{mylightred3}{HTML}{D8A9AA}
\definecolor{mylightred4}{HTML}{ECD4D5}
\definecolor{mylightblue1}{HTML}{5A7DA5}
\definecolor{mylightblue2}{HTML}{7D99BA}
\definecolor{mylightblue3}{HTML}{B3C3D7}
\definecolor{mylightblue4}{HTML}{D3DCE8}
\definecolor{mydarkgreen}{HTML}{3E4822}
\definecolor{mylightgreen1}{HTML}{828859}
\definecolor{mylightgreen2}{HTML}{9AA075}
\definecolor{mylightgreen3}{HTML}{B8BC96}
\definecolor{mylightgreen4}{HTML}{D4D4B8}
\definecolor{mylightgray1}{HTML}{6F6F6F}
\definecolor{mylightgray2}{HTML}{999999}
\definecolor{mylightgray3}{HTML}{B4B4B4}
\definecolor{mylightgray4}{HTML}{DCDCDC}

\IEEEoverridecommandlockouts

\safemath{\Herm}{\textnormal{H}}

\frefformat{vario}{\fancyreffiglabelprefix}{Fig.~#1}

\begin{document}

\bstctlcite{IEEEexample:BSTcontrol}

\title{
A 0.32\,$\text{mm}^\text{2}$ 100\,Mb/s 223\,mW ASIC in 22FDX\\
for Joint Jammer Mitigation, Channel Estimation, \\
and SIMO Data Detection 

}
\author{\IEEEauthorblockN{Jonas Elmiger\textsuperscript{*}\!, Fabian Stuber\textsuperscript{*}\!, Oscar Casta\~neda, Gian Marti, and Christoph Studer} \\[-0.3cm]
\thanks{\!\!\textsuperscript{*}Equal contribution. 
Contact author: O. Casta\~neda (e-mail: caoscar@ethz.ch)}
\IEEEauthorblockA{\emph{Department of Information Technology and Electrical Engineering, ETH Zurich, Switzerland}
\thanks{This work has received funding from the Swiss State Secretariat for Education, Research, and Innovation (SERI) under the SwissChips initiative. The authors thank GlobalFoundries for providing silicon fabrication through the 22FDX University Program. The authors also thank Darja Nonaca and J\'er\'emy Guichemerre for their assistance with backend design and testing.}
} 
}

\maketitle

\begin{abstract}
We present the first single-input multiple-output (SIMO) receiver ASIC that jointly performs jammer mitigation, 
channel estimation, and data detection.  
The ASIC implements a recent algorithm called siMultaneous mitigAtion, Estimation, 
and Detection (MAED). MAED mitigates smart jammers via spatial filtering 
using a nonlinear optimization problem that unifies jammer estimation and nulling, channel estimation, 
and data detection to achieve state-of-the-art error-rate performance under jamming. 
The design supports eight receive antennas 
and enables mitigation of smart jammers as well as of barrage jammers. 
The ASIC is fabricated in 22\,nm FD-SOI, has a core area of 0.32\,mm$^\text{2}$, and achieves a throughput of 100\,Mb/s at 223\,mW,
 thus delivering 3$\times$ higher per-user throughput and 4.5$\times$ higher area efficiency than the state-of-the-art jammer-resilient detector. 
\end{abstract}

\section{Introduction}\label{sec:intro}

In a jamming attack, a hostile interferer transmits interference 
signals to disrupt a wireless communication system. Safety-critical
communication systems must be able to mitigate such jamming attacks~\cite{pirayesh2022jamming}. 
The proliferation of multi-antenna receivers provides a promising avenue 
for jammer mitigation based on spatial filtering. However, spatial 
filtering requires accurate estimates of a jammer's spatial signature, which 
is difficult to acquire for smart jammers that try to avoid estimation. 
Nonlinear methods have recently been proposed for mitigating such smart 
jammers~\cite{marti2023jmd, marti2023maed}. Of these, only the SANDMAN algorithm 
 from~\cite{marti2023jmd} has so far been implemented in hardware~\cite{bucheli2024vlsi}, 
which is necessary to support the throughput requirements of modern wireless systems. 
However, SANDMAN suffers from suboptimal error-rate performance as it separates
channel estimation from jammer mitigation and data detection~\cite{vikalo2006efficient}. Moreover, its hardware efficiency 
is orders of magnitude below that of non-jammer-resilient data detectors~\cite{bucheli2024vlsi}. 

\textit{Contributions:} 
We present an application-specific integrated circuit (ASIC) implementation of the MAED (short for siMultaneous mitigAtion, Estimation, and Detection) algorithm from~\cite{marti2023maed} for an $8\times1$ single-input multiple-output (SIMO) system.
MAED is the first receiver ASIC 
that unifies jammer estimation and mitigation, channel estimation, and data detection.
Our design mitigates smart jammers and---by leveraging joint channel estimation and data detection (JED)---outperforms the only other jammer-resilient receiver implementation~\cite{bucheli2024vlsi} in terms of error-rate performance. In addition, our ASIC measurements demonstrate an improved area efficiency compared to~\cite{bucheli2024vlsi}.
Moreover, our design is the first silicon-proven ASIC for JED.\footnote{To our knowledge, the only other ASICs for JED are~\cite{castaneda16a,castaneda2017vlsi}, which are not fabricated in silicon, require preprocessing, and do not mitigate jammers.}

\section{Prerequisites} \label{sec:system}

\subsection{System Model}
 \label{sec:model}

\begin{figure}
\centering
\includegraphics[width=0.9\columnwidth]{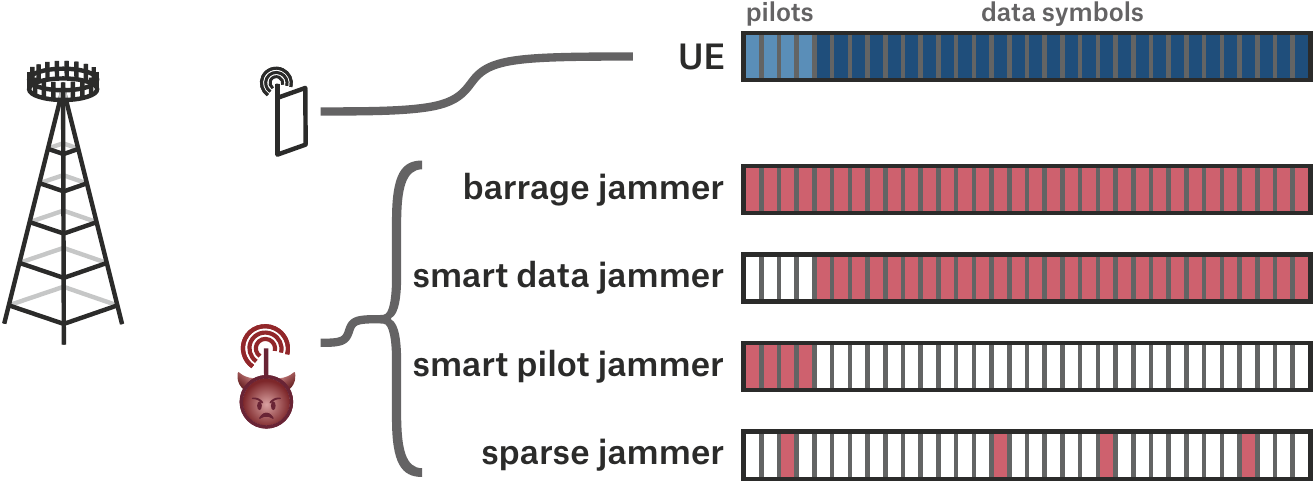}
\vspace{-1mm}
\caption{Considered scenario of a SIMO uplink attacked by a (smart) jammer.}
\vspace{-1mm}
\label{fig:system_setup}	
\end{figure}

We consider a SIMO system, which can model 
a single-antenna user equipment (UE) transmitting
to a $B$-antenna receiver under a potentially smart  
single-antenna jammer~(\fref{fig:system_setup}). 
We assume a flat-fading, block-fading channel with a coherence block of $K$ channel uses. The input-output relation is 
\begin{align}
	\bY = \bmh\tp{\bms} + \bmj\tp{\bmw} + \bN, \label{eq:io}
\end{align}
where $\bY\!\!\in\!\opC^{B\times K}$ is the receive signal,
$\bmh\!\in\!\opC^B$ and \mbox{$\bmj\!\in\!\opC^B$} are the UE and jammer channels, 
$\bms\in\opC^K$ and $\bmw\in\opC^K$ are the UE and jammer transmit signals, 
and $\bN\stackrel{\text{i.i.d.}}{\sim}\setC\setN(0,\No)$ models noise with per-entry variance~$\No$. 
The UE signal~\mbox{$\bms=\tp{[\tp{\bms_T},\tp{\bms_D}]}$} consists of 
$T$ pilots $\bms_T\in\opC^{T}$ and $D=K-T$ data symbols~\mbox{$\bms_D\in\setS^D$} from 
the QPSK constellation~$\setS$. %
Every entry~$s_i$ of $\bms$ satisfies $|s_i|^2=1$. 
We consider Rayleigh fading: $\bmh\stackrel{\text{i.i.d.}}{\sim}\setC\setN(0,1)$ and 
$\bmj\!\stackrel{\text{i.i.d.}}{\sim}\setC\setN(0,1)$.
Moreover, we assume that the jammer can be one of the following types 
(cf. \fref{fig:system_setup}):

\noindent
\textit{1) Barrage jammer:} 
This jammer sends $\bmw\stackrel{\text{i.i.d.}}{\sim}\setC\setN(0,\sigma^2)$. 

\noindent
\textit{2) Smart data jammer:}
This jammer sends $\bmw=\tp{[\tp{\bmw_T},\tp{\bmw_D}]}$ with $\bmw_T=\mathbf{0}\in\opC^T$ 
and $\bmw_D\stackrel{\text{i.i.d.}}{\sim}\setC\setN(0,\sigma^2)\in\opC^D$.

\noindent
\textit{3) Smart pilot jammer:}
This jammer sends $\bmw=\tp{[\tp{\bmw_T},\tp{\bmw_D}]}$ with 
$\bmw_T\stackrel{\text{i.i.d.}}{\sim}\setC\setN(0,\sigma^2)\in\opC^T$
and $\bmw_D=\mathbf{0}\in\opC^D$.

\noindent
\textit{4) Sparse jammer:}
This jammer sends $\bmw=\bmc\,\odot\,\bmz$, where~$\odot$ is the Hadamard product,
$\bmc$ is drawn from the distribution $\text{Unif}\big[\tilde\bmc\in\{0,1\}^K:\sum_k \tilde{c}_k=4\big]$,
and $\bmz\stackrel{\!\text{i.i.d.}\!}{\sim}\setC\setN(0,\sigma^2)$. 

The jammer power is characterized in terms of the receive jammer-to-signal ratio
$\rho\triangleq \frac{\|\bmj\tp{\bmw}\|_F^2}{\|\bmh\tp{\bms}\|_F^2}$, where $\|\bA\|_F$ is the Frobenius norm of a matrix $\bA$.

\subsection{MitigAtion, Estimation, and Detection (MAED)}
Smart jammers can be mitigated using the recently developed  MAED algorithm~\cite{marti2023maed}. 
MAED detects data in jammer-resilient fashion by approximately solving the optimization~problem
\begin{align}
	\min_{\tilde\bmj\in\opC^{B}, \tilde\bmh\in\opC^{B}, \tilde\bms_D\in\setS^D}
	\big\|\big(\bI_B-\tilde\bmj\pinv{\tilde\bmj}\big)\big(\bY - \tilde\bmh [\tp{\bms_T},\tp{\tilde\bms_D}] \big) \big\|_F^2, \label{eq:opt}
\end{align}
which unifies jammer estimation and mitigation (by solving for~$\tilde\bmj$), 
channel estimation (by solving for $\tilde\bmh$), and data detection (by solving for $\tilde\bms_D$). 
The problem in \eqref{eq:opt} can be transformed into a problem that only depends on $\tilde\bmj$ and $\tilde\bms_D$. 
It can then be solved approximately by alternating between power iterations in 
$\tilde\bmj$ and projected gradient descent steps in $\tilde\bms_D$. The resulting method is 
shown in \fref{alg:maed}. 
The proximal operator $\text{prox}(\tilde\bms_{(t)};\bms_T)$ on line 9 replaces the first $T$ entries of $\tilde\bms_{(t)}$ with~$\bms_T$ and clips the remaining $D$ entries to the convex hull of the constellation~$\setS$.

\begin{figure}[t]
\scalebox{0.9}{
\begin{minipage}[b]{0.63\columnwidth}

\tikz[remember picture,overlay] \fill[mylightgreen4, opacity=0.3] (0,-2.3) -- (5.4,-2.3) -- (5.6,-2.2) -- (9.8,-2.2) -- (9.8,-3.2) -- (5.6,-3.2) -- (5.4,-3.1) -- (0,-3.1) ;

\vspace{-4mm}
\tikz[remember picture,overlay] \fill[mylightgreen4, opacity=0.3] (0,-3.65) -- (5.4,-3.65) -- (5.6,-3.4) -- (9.8,-3.4) -- (9.8,-4.45) -- (5.6,-4.45) -- (5.4,-4.25) -- (0,-4.25) ;

\vspace{-4mm}
\tikz[remember picture,overlay] \fill[mylightgreen4, opacity=0.3] (0,-4.65) -- (5.4,-4.65) -- (5.6,-4.6) -- (9.8,-4.6) -- (9.8,-6) -- (5.6,-6) -- (5.4,-5.95) -- (0,-5.95) ;

\begin{algorithm}[H]
  \caption{SIMO MAED \cite{marti2023maed}}
  \label{alg:maed}
  \begin{algorithmic}[1]
	\setstretch{1.1}
    \State \textbf{input}: $\bY,\tau_{(0)},...,\tau_{(t_{\max}-1)},\bms_T$
    \State $\tilde\bms_{(0)} = \tp{[\tp{\bms_T},\tp{\mathbf{0}_{D\times1}}]}$
    \For{$t=0$ {\bfseries to} $t_{\max}-1$}
    	\vspace{0.25mm}
    	\State $\bE_{(t)} = \bY\Big(\bI_K - \frac{\tilde\bms_{(t)}^\ast\tp{\tilde\bms_{(t)}}}{\|\tilde\bms_{(t)}\|_2^2}\!\Big)$
    	\vspace{1.5mm}
    	\State $\bmu_{(t)} = \text{PRNG}(\,)$
    	\vspace{1.5mm}
    	\State $\tilde\bmj_{(t)} = \bE_{(t)}\herm{\bE_{(t)}}\bmu_{(t)}$
    	\vspace{5mm}
    	\State $\bP_{(t)} = \bI_B - \frac{\tilde\bmj_{(t)}\herm{\tilde\bmj_{(t)}}}{\|\tilde\bmj_{(t)}\|_2^2}$
    	\State $\nabla_{(t)} = -\frac{1}{\|\tilde\bms_{(t)}\|_2^2}
    	\tp{\tilde\bms_{(t)}}\herm{\bY}\bP_{(t)}\bE_{(t)}$
    	\vspace{2mm}
    	\State $\tilde\bms_{(t+1)} = \text{prox}\big(\tilde\bms_{(t)}\!-\tau_{(t)}\tp{\nabla_{(t)}};\bms_T\big)$
    \EndFor
    \State \textbf{output:} $\tilde\bms_{(t_{\max})}$
  \end{algorithmic}
\end{algorithm}
\end{minipage}

\hspace{-3mm}
\begin{minipage}[b]{0.46\columnwidth}
  \begin{algorithmic}[1]
	\setstretch{1.0}
  	\algrenewcommand{\alglinenumber}[1]{\footnotesize 4a:} 
	\State $\bmx_{(t)} \!=\! (1/\|\tilde\bms_{(t)}\|_2^2)\bY\tilde\bms_{(t)}^\ast$\!\!
  	\algrenewcommand{\alglinenumber}[1]{\footnotesize 4b:} 
	\State $\bE_{(t)} \!= \bY - \bmx_{(t)}\tp{\tilde\bms_{(t)}}$
	\vspace{3.25mm}
  	\algrenewcommand{\alglinenumber}[1]{\footnotesize 6a:} 
  	\State $\bmv_{(t)} \!= \herm{\bE_{(t)}}\bmu_{(t)}$
  	\algrenewcommand{\alglinenumber}[1]{\footnotesize 6b:} 
  	\State $\tilde\bmj_{(t)} \!= \bE_{(t)}\bmv_{(t)}$
	\vspace{2.5mm}	
  	\algrenewcommand{\alglinenumber}[1]{\footnotesize 7a:}
  	\State $\bmz_{(t)} \!= \!\bmx_{(t)} \!-\! \tilde\bmj_{(t)} \!\!\left(\frac{\herm{\tilde\bmj_{(t)}}\bmx_{(t)}}{\|\tilde\bmj_{(t)}\|_2^2}  \!\right)$\!\!\!
  	\algrenewcommand{\alglinenumber}[1]{\footnotesize 8a:}
	\State $-\tau_{(t)}\!\herm{\nabla_{(t)}}\!\!=\!\herm{\bE_{(t)}}\!(\tau_{(t)}\bmz_{(t)})$
	\vspace{14.1mm}
  \end{algorithmic}
  
\end{minipage}
}

\end{figure}

\subsection{Algorithm Rearrangements for Hardware Implementation}

To arrive at an efficient hardware implementation, we rearrange the MAED algorithm as outlined on the right side of \fref{alg:maed}.
These rearrangements avoid matrix-matrix products and instead perform sequences of matrix-vector products which reduce the number of multiplications required by \fref{alg:maed}.
Consider the computation of $\tilde\bmj_{(t)}$ on line 6 as an example: Computing first the matrix-matrix product $\bE_{(t)}\herm{\bE_{(t)}}$ and then the matrix-vector product $(\bE_{(t)}\herm{\bE_{(t)}})\bmu_{(t)}$ requires $(K+1)B^2$ complex-valued multiplications, whereas computing first the matrix-vector product on line 6a and then the one on line 6b requires only $2KB$ complex-valued multiplications.
Furthermore, while the original MAED algorithm~\cite{marti2023maed} applies the Barzilai-Borwein method to determine the step sizes $\tau_{(t)}$, we empirically tune these step sizes to powers of two. Doing so saves the computations of the Barzilai-Borwein method and simplifies multiplication by $\tau_{(t)}$ into a simple arithmetic shift.

\section{VLSI architecture} \label{sec:arch}

\begin{figure}[tp]
\centering
\includegraphics[width=\columnwidth]{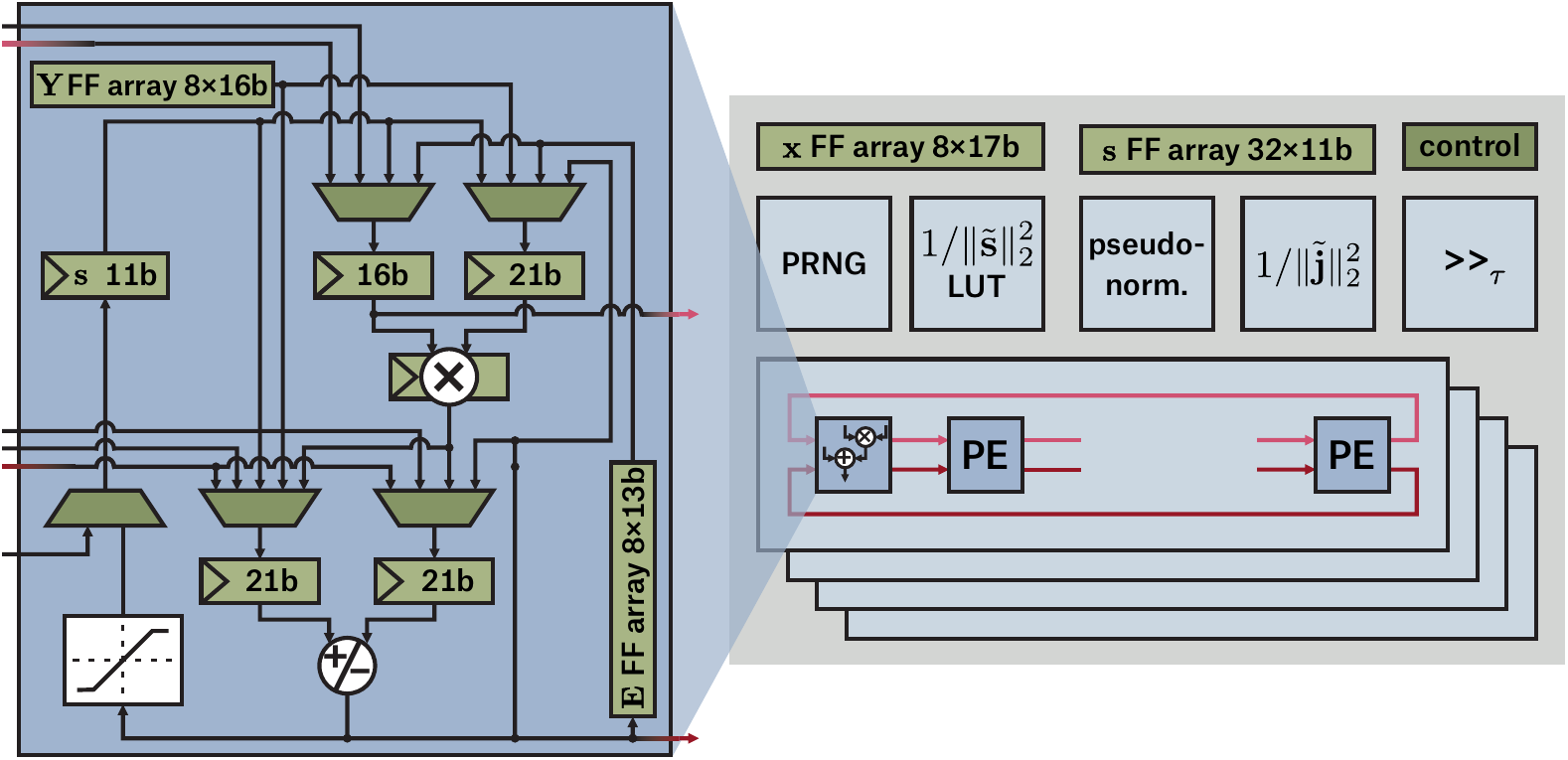}	
\vspace{-5mm}	
\caption{Top view of the MAED architecture with zoom-in on the architecture of the processing elements (PEs). The bitwidths shown for registers and flip-flop~(FF) arrays are per real and imaginary part of variables.}
\label{fig:arch}
\end{figure}

Our ASIC is designed to execute the rearranged \fref{alg:maed} for $B=8$, $K=32$ and $T=4$.
In what follows, we drop the iteration subscript ``$(t)$'' from the variables to simplify notation.
\fref{fig:arch} shows the top-level architecture, 
which consists of 32~processing elements (PEs) grouped in four slices
of 8~PEs each, as well as auxiliary modules for pseudorandom number generation~(PRNG), inversion, multiplication by the step size~$\tau$, and flip-flop (FF) arrays to store $\bmx$ and the output $\tilde\bms$.
\fref{fig:arch} also shows a PE's internals, 
which includes complex-valued multiply-accumulate circuitry, local FF arrays,
and a clipping unit for executing the $\text{prox}(\cdot)$ operator on line 9 of \fref{alg:maed}.

\subsection{Operation}

\begin{figure}[tp]
	\centering
    \vspace{-1mm}
    \subfigure[$\bA\bmx$]{
        \includegraphics[width=0.46\columnwidth]{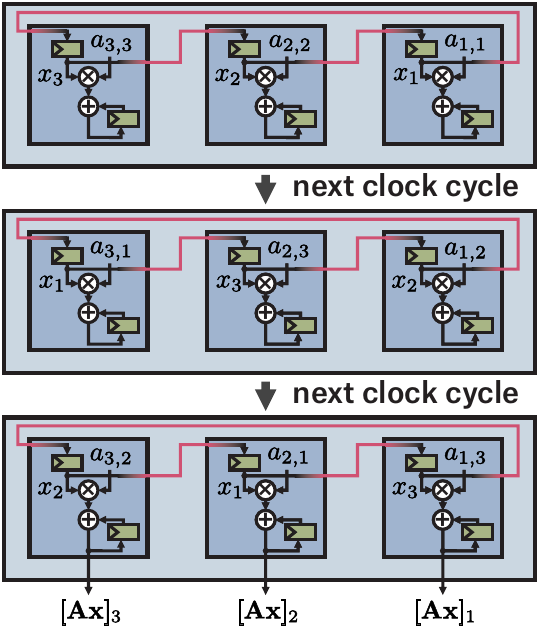}\label{fig:cannon1}
        }
	\hfill
	\subfigure[$\herm{\bA}\bmz$]{\
        \includegraphics[width=0.46\columnwidth]{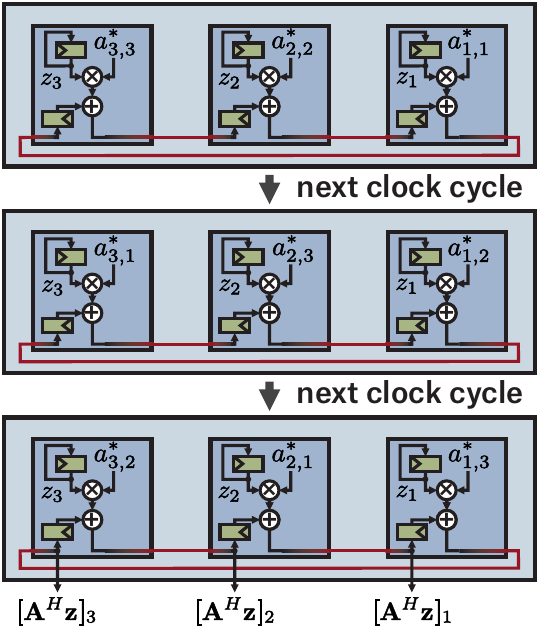}\label{fig:cannon2}
            }
    \vspace{-1mm}
    \caption{Cannon's algorithm and its Hermitian variant illustrated for a $3\times 3$ matrix-vector product: (a) To compute $\bA\bmx$, the $j$th PE stores $\bma_j$, the $j$th row of matrix $\bA$. Each PE starts with $x_k$ and multiplies it by $a_{j,k}$. The PEs then circularly exchange the entries of $\bmx$ and accumulate the partial products until completing $\bA\bmx$. (b) In the Hermitian variant, the PEs compute $\herm{\bA}\bmz$ by circularly exchanging the accumulated partial products while keeping the entries of $\bmz$ fixed. The entries of $\bma_j$ are read in the same order as for $\bA\bmx$. }\label{fig:cannon}
\end{figure}

\begin{figure}
\centering
    \hspace{-2mm}
    \subfigure[Pseudonormalization module]{
		\includegraphics[width=0.55\columnwidth]{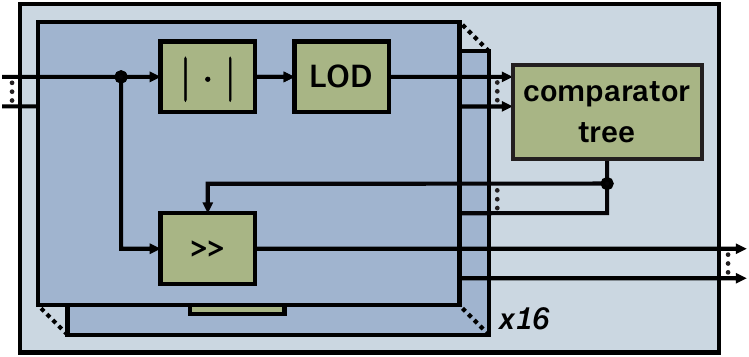}	\label{fig:pseudonormalization}
    }
    \hspace{-3mm}
    \subfigure[$\|\tilde\bmj\|^2_2$-inversion module]{
        \begin{minipage}[t]{0.42\columnwidth}
            \vspace{-2.313cm}
            \includegraphics[width=\linewidth]{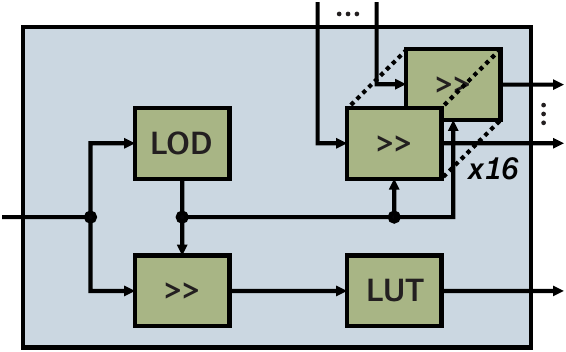}
        \end{minipage}
        \label{fig:subfig2}
    } 
    \hspace{-2mm}

~\vspace{-3mm}
\caption{Modules of the MAED architecture.}
\end{figure}

The  $8\times32$ matrix-vector product $\bY\tilde\bms^\ast$ on line 4a of \fref{alg:maed}
can be rewritten as $[\bY_1,\bY_2,\bY_3,\bY_4]\herm{[\tp{\tilde\bms_1},\tp{\tilde\bms_2},\tp{\tilde\bms_3},\tp{\tilde\bms_4}]}$, where $\bY_i\in\opC^{8\times 8}$ and $\tilde\bms_i\in\opC^8$, 
and is thus computed as the sum $\sum_i\bY_i\tilde\bms_i^\ast$ of four $8\times8$ matrix-vector products.
The $j$th PE of the $i$th slice stores the $j$th row of $\bY_i$, which enables the application of Cannon's algorithm~\cite{cannon69} (cf. \fref{fig:cannon1}) to compute the products $\bY_i\tilde\bms_i^\ast$ in parallel across the four PE slices.
The PE slices exchange their $\bY_i\tilde\bms_i^\ast$ to compute $\bY\tilde\bms^\ast$, which is stored across the PEs of the first slice.
The $8\times32$ matrix-vector product~$\bE\bmv$ on line~6b is computed by following the same procedure, 
which takes 13 clock cycles.
The matrix-vector products involving~$\herm{\bE}$ on lines 6a and 8a 
are each computed after 10 clock cycles by using a variant of Cannon's algorithm (cf. \fref{fig:cannon2}), which avoids rewriting memory to transpose $\bE$.

To complete line 4a, we need the reciprocal value of the 32-dimensional inner product  $\|\tilde\bms\|_2^2=\herm{\tilde\bms}\tilde\bms$.
The multipliers of the 32 PEs compute the partial products $\tilde s_i^\ast\tilde s_i$, which are then summed by reconfiguring the PEs' adders into an adder tree with five pipelined stages.
The resulting $\|\tilde\bms\|_2^2$ is confined to the interval $[4,32]$, and hence its inverse is simply computed via a look-up table (LUT).
The PEs of the first slice then multiply $1/\|\tilde\bms\|_2^2$ by $\bY\tilde\bms^\ast$ to obtain $\bmx$, which is stored in a FF array that is accessible to all PE slices.

Since the $j$th PE of the $i$th slice stores the $j$th row of $\bY_i$, it can compute line 4b by scaling $\tilde\bms_i$ by $x_j$. This is achieved in 12 clock cycles by circularly exchanging the entries of $\tilde\bms_i$ across the 8 PEs of the $i$th PE slice, just as it is done for Cannon's algorithm (cf. \fref{fig:cannon1}).
The next step, given by line~5, is to use the PRNG module to generate $\bmu$, which is used to start the power iteration that follows on lines 6a and 6b. The PRNG was implemented using xorshift~\cite{marsaglia2003xorshift} with a 64-bit state.

The vector $\tilde\bmj$ of line 6b corresponds to an estimate of the jammer channel $\bmj$, up to an unknown scale factor proportional to the jammer power. 
Since the jammer might be jamming at very high power (e.g., the received jammer power could exceed that of the UE by $\rho=30$\,dB), a large bitwidth is required to represent the entries of  $\tilde\bmj$, leading to an even larger bitwidth to represent $\|\tilde\bmj\|^2_2$.
In fact, the jammer-resilient detector from~\cite{bucheli2024vlsi} uses dedicated PEs with extended bitwidths to compute an analogous quantity.
We avoid this significant hardware overhead by noting that the subsequent algorithm operation (line 7a) is invariant to the scale of $\tilde\bmj$.
This allows us to first rescale $\tilde\bmj$ with the pseudonormalization module (cf. \fref{sec:pseudonorm}) to focus only on the bits that are relevant for the computation of $\|\tilde\bmj\|^2_2$.

The rescaled $\tilde\bmj$ is then used to compute the 8-dimensional inner products $\herm{\tilde\bmj}\bmx$ and $\|\tilde\bmj\|^2_2=\herm{\tilde\bmj}\tilde\bmj$ (line 7a). These inner products are each computed in 5 clock cycles using one PE slice. The $\|\tilde\bmj\|^2_2$-inversion module (described in \fref{sec:inversion}) then computes $1/\|\tilde\bmj\|^2_2$, which is multiplied by $\herm{\tilde\bmj}\bmx$. Using these quantities, the first PE slice computes $\bmz$ (line 7a) and scales its entries by the step size $\tau$ through the $\mathtt{>>}_\tau$ arithmetic shifters.
The resulting $\tau\bmz$ is broadcasted to all PE slices in order to compute $\herm{\bE}(\tau\bmz)$ (line 8a) using the Hermitian variant of Cannon's algorithm (cf. \fref{fig:cannon2}).
Finally, each PE performs the gradient descent step and proximal operation for one entry of $\tilde\bms$ to obtain the new iterate (line 9).
This completes one iteration of the MAED algorithm after 83 clock cycles.

\subsection{Pseudonormalization Module}\label{sec:pseudonorm}
Na\"ively, the potential dynamic range of $\tilde\bmj$ would require large 
bitwidths to faithfully represent $\tilde\bmj$ on line 7a
(since it is not known \emph{a priori} where the relevant bits of $\tilde\bmj$ are located). 
However, since line 7a is invariant to the scale of $\tilde\bmj$,
the pseudonormalization solves this issue by rescaling $\tilde\bmj$
such that $\|\tilde\bmj\|_2^2\in[1,32)$, which enables subsequent operations to discard fractional bits not needed for the required computational precision. 
Our module finds the largest (in magnitude) entry~$j_\textnormal{max}$  across the real and imaginary parts of $\tilde\bmj$ using leading one detectors (LODs), and divides all entries by $2^{\lfloor\log_2(j_\textnormal{max})\rfloor}$.

\subsection{$\|\tilde\bmj\|_2^2$-Inversion Module}\label{sec:inversion}
The $\|\tilde\bmj\|_2^2$-inversion module (cf. \fref{fig:subfig2}) starts by scaling $\|\tilde\bmj\|_2^2$ to be within the interval $[1,2)$, to then compute its scaled inverse with a LUT. To obtain the desired $1/\|\tilde\bmj\|_2^2$, the result from the LUT should be scaled back, but this would require an increased bitwidth at the input of the multiplier computing $(\herm{\tilde\bmj}\bmx)/\|\tilde\bmj\|_2^2$ to not lose the LUT's precision. Instead, we scale $\bmx$ when computing $\herm{\tilde\bmj}\bmx$, so that both $\herm{\tilde\bmj}\bmx$ and $\|\tilde\bmj\|_2^2$ are scaled by the same factor. While this entails some loss in the precision of the entries of $\bmx$ for computing $\herm{\tilde\bmj}\bmx$, our results demonstrate that the performance is not affected (cf. \fref{sec:ber}).

\section{ASIC Implementation Results}\label{sec:res}

\subsection{Bit Error-Rate (BER) Performance} \label{sec:ber}

\begin{figure}[tbp]
	\centering
    \vspace{-1mm}
    \subfigure[barrage jammer]{
        \includegraphics[width=0.465\columnwidth]{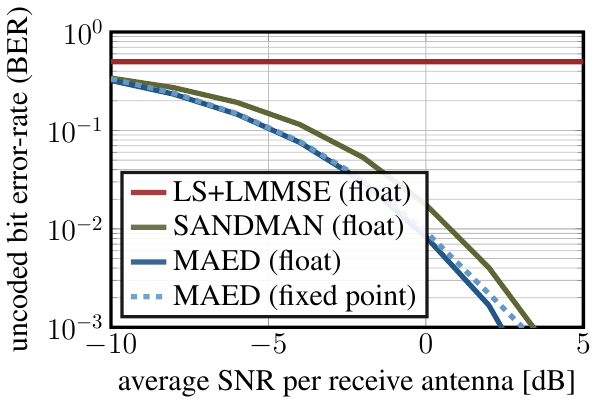}\label{fig:barrage}
        }
	\hfill
	\subfigure[smart data jammer]{\
        \includegraphics[width=0.465\columnwidth]{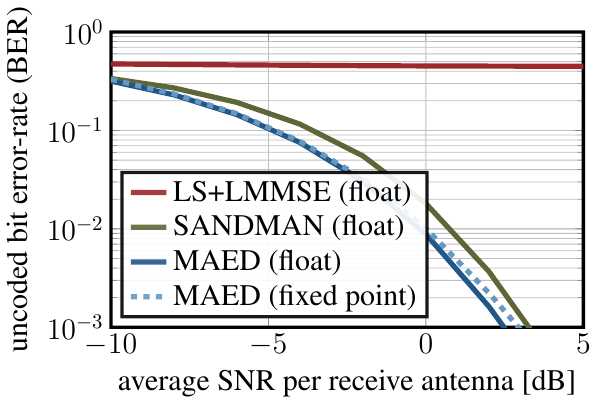}\label{fig:data}
            }
    \subfigure[smart pilot jammer]{
        \includegraphics[width=0.465\columnwidth]{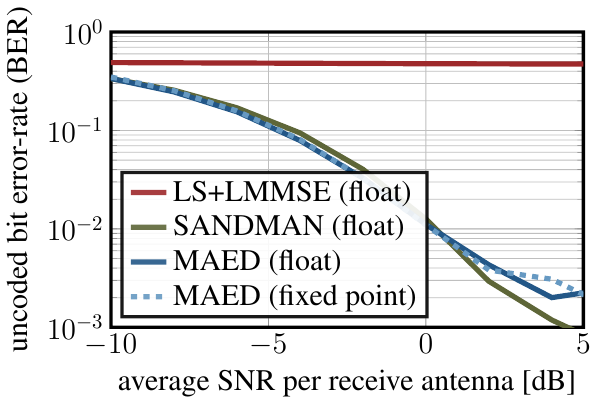}\label{fig:pilot}
        }
	\hfill
	\subfigure[sparse jammer]{\
        \includegraphics[width=0.465\columnwidth]{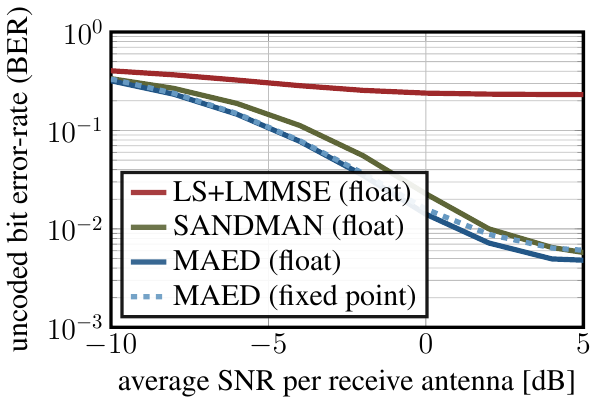}\label{fig:sparse}
            }
     \vspace{-2mm}
    \caption{Uncoded bit error-rate (BER) of different SIMO receivers as a function of the SNR for different jamming scenarios (cf. \fref{sec:model}). The receive jammer-to-signal ratio of all jammers is equal to $\rho=30$\,dB.} \label{fig:bers}
\end{figure}

\fref{fig:bers} shows the uncoded bit error-rate (BER) of different SIMO receivers that detect QPSK signals while under attack from one of four different jammers (cf. \fref{sec:model}).
We vary the signal-to-noise ratio (SNR) per receive antenna, which is defined as $\text{SNR}=\|\bmh\|_2^2/(B\No)$, whereas all jammers interfere with a jammer-to-signal ratio of $\rho=30$\,dB. 
The compared receivers are a non-mitigating receiver with least squares (LS) channel estimation 
and linear minimum mean squared error (LMMSE) data detection as in~\cite{prabhu20173}, an adapted version of the jammer-mitigating SANDMAN method~\cite{marti2023jmd, bucheli2024vlsi} for SIMO,\footnote{
SANDMAN resembles MAED in that it also mitigates the jammer jointly with detecting the receive data, 
and thus is able to mitigate smart jammers as well as barrage jammers. However, SANDMAN 
estimates the UE channel separately and thus fails to harvest the gains associated with JED
\cite{vikalo2006efficient, castaneda2017vlsi}.}
and the MAED algorithm implemented in this work. 

\fref{fig:bers} shows that the BER of the non-mitigating LS+LMMSE receiver does not fall below 10\% 
for any of the jammer types. In contrast, both SANDMAN and MAED succeed in mitigating all jammers and 
reach BERs below 1\% at high SNR. However, MAED outperforms SANDMAN 
thanks to the gains afforded by JED. 
Finally, the BER curves of the fixed-point implementation of MAED closely follow those of the double-precision floating-point reference, which demonstrates that the selected numerical representation scheme of the silicon implementation is sufficiently precise for practical purposes. 

\subsection{ASIC Measurements and Comparison}

\begin{figure}[tbp]
\centering
\includegraphics[width=0.95\columnwidth]{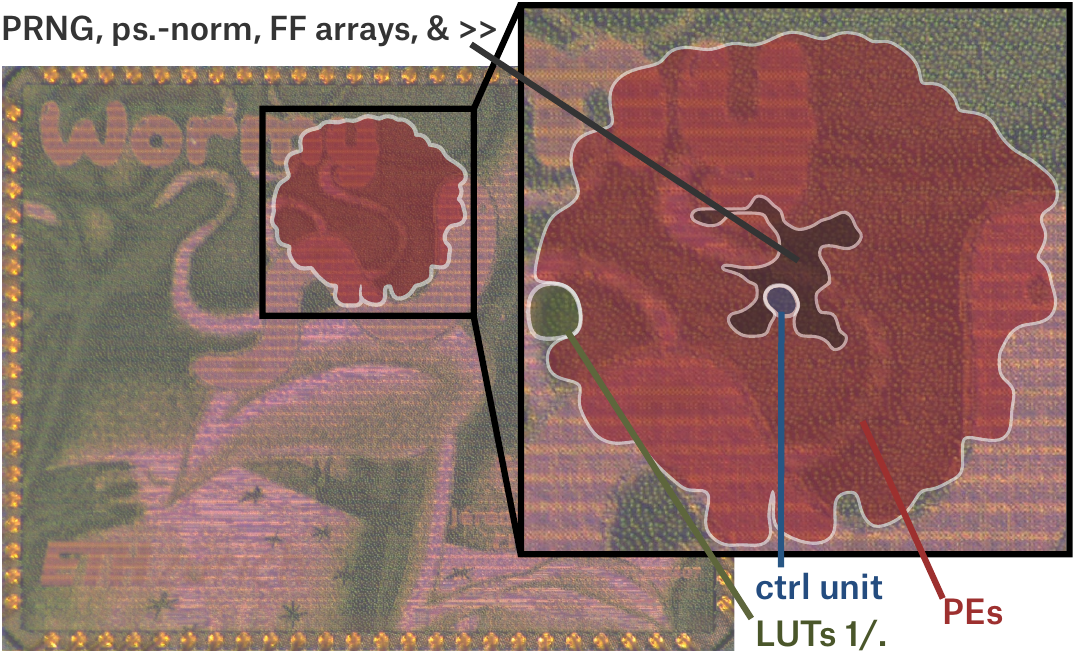}
\vspace{-3mm}
\caption{Micrograph of the $2.5$\,mm $\times$ $2$\,mm ASIC with the area containing the MAED design zoomed-in; the rest of the ASIC contains unrelated designs.}
\label{fig:amoeba}	
\end{figure}

\begin{figure}[tbp]
	\centering
	\hspace{-2mm}
    \subfigure{
        \includegraphics[height=3.95cm]{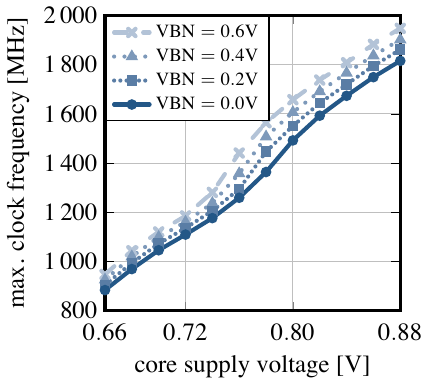}\label{fig:frequency}
        }
	\hspace{-5mm}
	\subfigure{\
        \includegraphics[height=3.95cm]{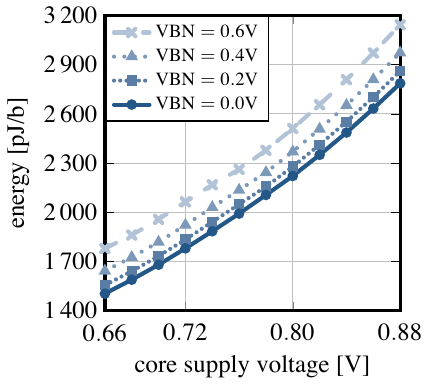}\label{fig:energy}
            }
	\hspace{-2mm}
	
	\vspace{-2mm}
    \caption{Measured maximum clock frequency and energy per bit as the core supply and NMOS forward body biasing (VBN) vary; VBP is held at $0$\,V.} \label{fig:measurements}
\end{figure}

\fref{fig:amoeba} shows the micrograph of the $5$\,mm$^\text{2}$ 22\,nm FD-SOI ASIC that contains our design along with other, unrelated designs. Our design occupies a core area of $0.32$\,mm$^\text{2}$. 
At $0.8$\,V nominal core supply, zero body biasing, and $300$\,K room temperature, our design achieves
a maximum clock frequency of $1.49$\,GHz, which, for $t_\textnormal{max}=10$ iterations, corresponds to $100$\,Mb/s throughput, 
$314$\,Mb/s/mm$^\text{2}$ area efficiency, and $2.2$\,nJ/b energy when facing a $\rho=30$\,dB jammer.
\fref{fig:measurements} shows the maximum clock frequency and energy
as a function of the core supply voltage, with and without forward body biasing. 

MAED is the first jammer-resilient JED ASIC, so a direct comparison to other detectors is not entirely fair. 
Table \ref{tbl:comp} shows such a comparison nonetheless. 
Compared to the only other jammer-mitigating detector in the literature~\cite{bucheli2024vlsi}, MAED achieves 3$\times$ higher per-user throughput and 4.5$\times$ higher area efficiency at the same energy efficiency. 
Compared to the most efficient SIMO-JED detector in the literature~\cite{castaneda2017vlsi}, MAED exhibits orders of magnitude lower efficiency, which is the cost of mitigating smart jammers; see also the efficiency difference between the \mbox{(non-)jammer}-resilient MIMO detectors \cite{prabhu20173} and \cite{bucheli2024vlsi}.

\begin{table}[tbp]
\setlength{\tabcolsep}{4pt} %
    \caption{Measurement Results and ASIC comparison}
    \label{tbl:comp}
    \vspace{-0.2cm}
    \centering
    \scalebox{0.8}{
\begin{tabular}{@{} l c | c c c @{}} 
\toprule[0.2em]
\multirow{2}{4em}{}
& 
This work\!  & Bucheli \cite{bucheli2024vlsi} & \!\!Casta\~neda \cite{castaneda2017vlsi}\!\! & Prabhu~\cite{prabhu20173} \\
\midrule[0.15em]
Max. receiver antennas	 	& 8 & 32  & 128 & 128 \\ 
Max. UEs  			& 1 & 8  & 1 & 8 \\ 
Modulation [QAM]\!\!\!\!	&  4 &  16 &  4  &  256 \\
Algorithm 
					& MAED & SANDMAN & PrOX & LMMSE \\
Jammer mitigation\!\!
 					& {\bf yes}  & {\bf yes} & {\bf no} & {\bf no}\\ 
Joint est.\ \& det.\ (JED)\!
					& {\bf yes} & {\bf no} & {\bf yes}$^a$ & {\bf no} \\
\midrule[0.15em]
Technology [nm] 	& 22  & 22 & 40 & 28 \\
Core supply [V]        & 0.8 & 0.78 & 1.1 & 0.9 \\
Core area~[$\text{mm}^2$] 	& 0.32 & 3.78 & 0.30 & $-$ \\
Max. frequency~[MHz] 			& 1\,492 & 320 & 695 & 300 \\
Throughput~[Mb/s]\!\! 
						& 100 & 267 & 412 & 300 \\ 
Power~[mW] 				& 223 & 583 & 58 & 18 \\
Area~eff.$^b$~[Mb/s/$\text{mm}^2$]\!\!\!\!\! 
						&  314 & 70 & 8\,150 & $-$ \\
Energy$^b$~[pJ/b]		& 2\,218 & 2\,300 & 41 & 37 \\
\bottomrule[0.2em]
\end{tabular}}\\[0.1cm]
\raggedright  \phantom{-}{\scriptsize$^\textit{a}$Not validated in silicon, $^\textit{b}$technology normalized to 22\,nm at $0.8$\,V core supply.}
\end{table}

\section{Conclusions}
We have presented the \emph{first} ASIC for joint data detection, jammer mitigation, and channel estimation in SIMO, which is also the first silicon-proven ASIC for JED.
Our design outperforms the only jammer-resilient detector in the open literature in terms of error-rate performance, \mbox{per-user} throughput, and area efficiency, without degrading energy efficiency. Jammer resilience comes at a steep premium, but without it, communication in hostile environments is doomed to fail.

\vfill

\linespread{1}

\end{document}

%% file: macros/vmr-symbols-vecbold.tex
\usepackage{amssymb}
\usepackage{amsfonts}
\usepackage{mathrsfs}
\usepackage{xspace}
\usepackage{bm}
\usepackage{upgreek}

\newcommand{\safemath}[2]{\newcommand{#1}{\ensuremath{#2}\xspace}}

\safemath{\bma}{\mathbf{a}}
\safemath{\bmb}{\mathbf{b}}
\safemath{\bmc}{\mathbf{c}}
\safemath{\bmd}{\mathbf{d}}
\safemath{\bme}{\mathbf{e}}
\safemath{\bmf}{\mathbf{f}}
\safemath{\bmg}{\mathbf{g}}
\safemath{\bmh}{\mathbf{h}}
\safemath{\bmi}{\mathbf{i}}
\safemath{\bmj}{\mathbf{j}}
\safemath{\bmk}{\mathbf{k}}
\safemath{\bml}{\mathbf{l}}
\safemath{\bmm}{\mathbf{m}}
\safemath{\bmn}{\mathbf{n}}
\safemath{\bmo}{\mathbf{o}}
\safemath{\bmp}{\mathbf{p}}
\safemath{\bmq}{\mathbf{q}}
\safemath{\bmr}{\mathbf{r}}
\safemath{\bms}{\mathbf{s}}
\safemath{\bmt}{\mathbf{t}}
\safemath{\bmu}{\mathbf{u}}
\safemath{\bmv}{\mathbf{v}}
\safemath{\bmw}{\mathbf{w}}
\safemath{\bmx}{\mathbf{x}}
\safemath{\bmy}{\mathbf{y}}
\safemath{\bmz}{\mathbf{z}}
\safemath{\bmzero}{\mathbf{0}}
\safemath{\bmone}{\mathbf{1}}

\bmdefine{\biad}{a}
\bmdefine{\bibd}{b}
\bmdefine{\bicd}{c}
\bmdefine{\bidd}{d}
\bmdefine{\bied}{e}
\bmdefine{\bifd}{f}
\bmdefine{\bigd}{g}
\bmdefine{\bihd}{h}
\bmdefine{\biid}{i}
\bmdefine{\bijd}{j}
\bmdefine{\bikd}{k}
\bmdefine{\bild}{l}
\bmdefine{\bimd}{m}
\bmdefine{\bind}{n}
\bmdefine{\biod}{o}
\bmdefine{\bipd}{p}
\bmdefine{\biqd}{q}
\bmdefine{\bird}{r}
\bmdefine{\bisd}{s}
\bmdefine{\bitd}{t}
\bmdefine{\biud}{u}
\bmdefine{\bivd}{v}
\bmdefine{\biwd}{w}
\bmdefine{\bixd}{x}
\bmdefine{\biyd}{y}
\bmdefine{\bizd}{z}

\bmdefine{\bixid}{\xi}
\bmdefine{\bilambdad}{\lambda}
\bmdefine{\bimud}{\mu}
\bmdefine{\bithetad}{\theta}
\bmdefine{\biphid}{\phi}
\bmdefine{\bideltad}{\delta}

\safemath{\bmia}{\biad}
\safemath{\bmib}{\bibd}
\safemath{\bmic}{\bicd}
\safemath{\bmid}{\bidd}
\safemath{\bmie}{\bied}
\safemath{\bmif}{\bifd}
\safemath{\bmig}{\bigd}
\safemath{\bmih}{\bihd}
\safemath{\bmii}{\biid}
\safemath{\bmij}{\bijd}
\safemath{\bmik}{\bikd}
\safemath{\bmil}{\bild}
\safemath{\bmim}{\bimd}
\safemath{\bmin}{\bind}
\safemath{\bmio}{\biod}
\safemath{\bmip}{\bipd}
\safemath{\bmiq}{\biqd}
\safemath{\bmir}{\bird}
\safemath{\bmis}{\bisd}
\safemath{\bmit}{\bitd}
\safemath{\bmiu}{\biud}
\safemath{\bmiv}{\bivd}
\safemath{\bmiw}{\biwd}
\safemath{\bmix}{\bixd}
\safemath{\bmiy}{\biyd}
\safemath{\bmiz}{\bizd}

\safemath{\bmxi}{\bixid}
\safemath{\bmlambda}{\bilambdad}
\safemath{\bmmu}{\bimud}
\safemath{\bmtheta}{\bithetad}
\safemath{\bmphi}{\biphid}
\safemath{\bmdelta}{\bideltad}

\safemath{\bA}{\mathbf{A}}
\safemath{\bB}{\mathbf{B}}
\safemath{\bC}{\mathbf{C}}
\safemath{\bD}{\mathbf{D}}
\safemath{\bE}{\mathbf{E}}
\safemath{\bF}{\mathbf{F}}
\safemath{\bG}{\mathbf{G}}
\safemath{\bH}{\mathbf{H}}
\safemath{\bI}{\mathbf{I}}
\safemath{\bJ}{\mathbf{J}}
\safemath{\bK}{\mathbf{K}}
\safemath{\bL}{\mathbf{L}}
\safemath{\bM}{\mathbf{M}}
\safemath{\bN}{\mathbf{N}}
\safemath{\bO}{\mathbf{O}}
\safemath{\bP}{\mathbf{P}}
\safemath{\bQ}{\mathbf{Q}}
\safemath{\bR}{\mathbf{R}}
\safemath{\bS}{\mathbf{S}}
\safemath{\bT}{\mathbf{T}}
\safemath{\bU}{\mathbf{U}}
\safemath{\bV}{\mathbf{V}}
\safemath{\bW}{\mathbf{W}}
\safemath{\bX}{\mathbf{X}}
\safemath{\bY}{\mathbf{Y}}
\safemath{\bZ}{\mathbf{Z}}

\safemath{\bZero}{\mathbf{0}}
\safemath{\bOne}{\mathbf{1}}
\safemath{\bDelta}{\mathbf{\Delta}}
\safemath{\bLambda}{\mathbf{\UpLambda}}
\safemath{\bPhi}{\mathbf{\Upphi}}
\safemath{\bSigma}{\mathbf{\Upsigma}}
\safemath{\bOmega}{\mathbf{\Upomega}}
\safemath{\bTheta}{\mathbf{\Uptheta}}

\bmdefine{\biAd}{A}
\bmdefine{\biBd}{B}
\bmdefine{\biCd}{C}
\bmdefine{\biDd}{D}
\bmdefine{\biEd}{E}
\bmdefine{\biFd}{F}
\bmdefine{\biGd}{G}
\bmdefine{\biHd}{H}
\bmdefine{\biId}{I}
\bmdefine{\biJd}{J}
\bmdefine{\biKd}{K}
\bmdefine{\biLd}{L}
\bmdefine{\biMd}{M}
\bmdefine{\biOd}{N}
\bmdefine{\biPd}{O}
\bmdefine{\biQd}{P}
\bmdefine{\biRd}{R}
\bmdefine{\biSd}{S}
\bmdefine{\biTd}{T}
\bmdefine{\biUd}{U}
\bmdefine{\biVd}{V}
\bmdefine{\biWd}{W}
\bmdefine{\biXd}{X}
\bmdefine{\biYd}{Y}
\bmdefine{\biZd}{Z}

\bmdefine{\biDelta}{\Delta}
\bmdefine{\biLambda}{\Lambda}
\bmdefine{\biPhi}{\Phi}
\bmdefine{\biSigma}{\Sigma}
\bmdefine{\biOmega}{\Omega}
\bmdefine{\biTheta}{\Theta}

\safemath{\bimA}{\biAd}
\safemath{\bimB}{\biBd}
\safemath{\bimC}{\biCd}
\safemath{\bimD}{\biDd}
\safemath{\bimE}{\biEd}
\safemath{\bimF}{\biFd}
\safemath{\bimG}{\biGd}
\safemath{\bimH}{\biHd}
\safemath{\bimI}{\biId}
\safemath{\bimJ}{\biJd}
\safemath{\bimK}{\biKd}
\safemath{\bimL}{\biLd}
\safemath{\bimM}{\biMd}
\safemath{\bimN}{\biNd}
\safemath{\bimO}{\biOd}
\safemath{\bimP}{\biPd}
\safemath{\bimQ}{\biQd}
\safemath{\bimR}{\biRd}
\safemath{\bimS}{\biSd}
\safemath{\bimT}{\biTd}
\safemath{\bimU}{\biUd}
\safemath{\bimV}{\biVd}
\safemath{\bimW}{\biWd}
\safemath{\bimX}{\biXd}
\safemath{\bimY}{\biYd}
\safemath{\bimZ}{\biZd}

\safemath{\bimDelta}{\biDelta}
\safemath{\bimLambda}{\biLambda}
\safemath{\bimPhi}{\biPhi}
\safemath{\bimSigma}{\biSigma}
\safemath{\bimOmega}{\biOmega}
\safemath{\bimTheta}{\biTheta}

\safemath{\setA}{\mathcal{A}}
\safemath{\setB}{\mathcal{B}}
\safemath{\setC}{\mathcal{C}}
\safemath{\setD}{\mathcal{D}}
\safemath{\setE}{\mathcal{E}}
\safemath{\setF}{\mathcal{F}}
\safemath{\setG}{\mathcal{G}}
\safemath{\setH}{\mathcal{H}}
\safemath{\setI}{\mathcal{I}}
\safemath{\setJ}{\mathcal{J}}
\safemath{\setK}{\mathcal{K}}
\safemath{\setL}{\mathcal{L}}
\safemath{\setM}{\mathcal{M}}
\safemath{\setN}{\mathcal{N}}
\safemath{\setO}{\mathcal{O}}
\safemath{\setP}{\mathcal{P}}
\safemath{\setQ}{\mathcal{Q}}
\safemath{\setR}{\mathcal{R}}
\safemath{\setS}{\mathcal{S}}
\safemath{\setT}{\mathcal{T}}
\safemath{\setU}{\mathcal{U}}
\safemath{\setV}{\mathcal{V}}
\safemath{\setW}{\mathcal{W}}
\safemath{\setX}{\mathcal{X}}
\safemath{\setY}{\mathcal{Y}}
\safemath{\setZ}{\mathcal{Z}}
\safemath{\emptySet}{\varnothing}

\safemath{\colA}{\mathscr{A}}
\safemath{\colB}{\mathscr{B}}
\safemath{\colC}{\mathscr{C}}
\safemath{\colD}{\mathscr{D}}
\safemath{\colE}{\mathscr{E}}
\safemath{\colF}{\mathscr{F}}
\safemath{\colG}{\mathscr{G}}
\safemath{\colH}{\mathscr{H}}
\safemath{\colI}{\mathscr{I}}
\safemath{\colJ}{\mathscr{J}}
\safemath{\colK}{\mathscr{K}}
\safemath{\colL}{\mathscr{L}}
\safemath{\colM}{\mathscr{M}}
\safemath{\colN}{\mathscr{N}}
\safemath{\colO}{\mathscr{O}}
\safemath{\colP}{\mathscr{P}}
\safemath{\colQ}{\mathscr{Q}}
\safemath{\colR}{\mathscr{R}}
\safemath{\colS}{\mathscr{S}}
\safemath{\colT}{\mathscr{T}}
\safemath{\colU}{\mathscr{U}}
\safemath{\colV}{\mathscr{V}}
\safemath{\colW}{\mathscr{W}}
\safemath{\colX}{\mathscr{X}}
\safemath{\colY}{\mathscr{Y}}
\safemath{\colZ}{\mathscr{Z}}

\safemath{\opA}{\mathbb{A}}
\safemath{\opB}{\mathbb{B}}
\safemath{\opC}{\mathbb{C}}
\safemath{\opD}{\mathbb{D}}
\safemath{\opE}{\mathbb{E}}
\safemath{\opF}{\mathbb{F}}
\safemath{\opG}{\mathbb{G}}
\safemath{\opH}{\mathbb{H}}
\safemath{\opI}{\mathbb{I}}
\safemath{\opJ}{\mathbb{J}}
\safemath{\opK}{\mathbb{K}}
\safemath{\opL}{\mathbb{L}}
\safemath{\opM}{\mathbb{M}}
\safemath{\opN}{\mathbb{N}}
\safemath{\opO}{\mathbb{O}}
\safemath{\opP}{\mathbb{P}}
\safemath{\opQ}{\mathbb{Q}}
\safemath{\opR}{\mathbb{R}}
\safemath{\opS}{\mathbb{S}}
\safemath{\opT}{\mathbb{T}}
\safemath{\opU}{\mathbb{U}}
\safemath{\opV}{\mathbb{V}}
\safemath{\opW}{\mathbb{W}}
\safemath{\opX}{\mathbb{X}}
\safemath{\opY}{\mathbb{Y}}
\safemath{\opZ}{\mathbb{Z}}
\safemath{\opZero}{\mathbb{O}}
\safemath{\identityop}{\opI}

\safemath{\veca}{\bma}
\safemath{\vecb}{\bmb}
\safemath{\vecc}{\bmc}
\safemath{\vecd}{\bmd}
\safemath{\vece}{\bme}
\safemath{\vecf}{\bmf}
\safemath{\vecg}{\bmg}
\safemath{\vech}{\bmh}
\safemath{\veci}{\bmi}
\safemath{\vecj}{\bmj}
\safemath{\veck}{\bmk}
\safemath{\vecl}{\bml}
\safemath{\vecm}{\bmm}
\safemath{\vecn}{\bmn}
\safemath{\veco}{\bmo}
\safemath{\vecp}{\bmp}
\safemath{\vecq}{\bmq}
\safemath{\vecr}{\bmr}
\safemath{\vecs}{\bms}
\safemath{\vect}{\bmt}
\safemath{\vecu}{\bmu}
\safemath{\vecv}{\bmv}
\safemath{\vecw}{\bmw}
\safemath{\vecx}{\bmx}
\safemath{\vecy}{\bmy}
\safemath{\vecz}{\bmz}

\safemath{\veczero}{\bmzero}
\safemath{\vecone}{\bmone}
\safemath{\vecxi}{\bmxi}
\safemath{\veclambda}{\bmlambda}
\safemath{\vecmu}{\bmmu}
\safemath{\vectheta}{\bmtheta}
\safemath{\vecphi}{\bmphi}
\safemath{\vecdelta}{\bmdelta}

\safemath{\matA}{\bA}
\safemath{\matB}{\bB}
\safemath{\matC}{\bC}
\safemath{\matD}{\bD}
\safemath{\matE}{\bE}
\safemath{\matF}{\bF}
\safemath{\matG}{\bG}
\safemath{\matH}{\bH}
\safemath{\matI}{\bI}
\safemath{\matJ}{\bJ}
\safemath{\matK}{\bK}
\safemath{\matL}{\bL}
\safemath{\matM}{\bM}
\safemath{\matN}{\bN}
\safemath{\matO}{\bO}
\safemath{\matP}{\bP}
\safemath{\matQ}{\bQ}
\safemath{\matR}{\bR}
\safemath{\matS}{\bS}
\safemath{\matT}{\bT}
\safemath{\matU}{\bU}
\safemath{\matV}{\bV}
\safemath{\matW}{\bW}
\safemath{\matX}{\bX}
\safemath{\matY}{\bY}
\safemath{\matZ}{\bZ}
\safemath{\matzero}{\bmzero}

\safemath{\matDelta}{\bDelta}
\safemath{\matLambda}{\bLambda}
\safemath{\matPhi}{\bPhi}
\safemath{\matSigma}{\bSigma}
\safemath{\matOmega}{\bOmega}
\safemath{\matTheta}{\bTheta}

\safemath{\matidentity}{\matI}
\safemath{\matone}{\matO}

\safemath{\rnda}{A}
\safemath{\rndb}{B}
\safemath{\rndc}{C}
\safemath{\rndd}{D}
\safemath{\rnde}{E}
\safemath{\rndf}{F}
\safemath{\rndg}{G}
\safemath{\rndh}{H}
\safemath{\rndi}{I}
\safemath{\rndj}{J}
\safemath{\rndk}{K}
\safemath{\rndl}{L}
\safemath{\rndm}{M}
\safemath{\rndn}{N}
\safemath{\rndo}{O}
\safemath{\rndp}{P}
\safemath{\rndq}{Q}
\safemath{\rndr}{R}
\safemath{\rnds}{S}
\safemath{\rndt}{T}
\safemath{\rndu}{U}
\safemath{\rndv}{V}
\safemath{\rndw}{W}
\safemath{\rndx}{X}
\safemath{\rndy}{Y}
\safemath{\rndz}{Z}

\safemath{\rveca}{\bimA}
\safemath{\rvecb}{\bimB}
\safemath{\rvecc}{\bimC}
\safemath{\rvecd}{\bimD}
\safemath{\rvece}{\bimE}
\safemath{\rvecf}{\bimF}
\safemath{\rvecg}{\bimG}
\safemath{\rvech}{\bimH}
\safemath{\rveci}{\bimI}
\safemath{\rvecj}{\bimJ}
\safemath{\rveck}{\bimK}
\safemath{\rvecl}{\bimL}
\safemath{\rvecm}{\bimM}
\safemath{\rvecn}{\bimN}
\safemath{\rveco}{\bomO}
\safemath{\rvecp}{\bimP}
\safemath{\rvecq}{\bimQ}
\safemath{\rvecr}{\bimR}
\safemath{\rvecs}{\bimS}
\safemath{\rvect}{\bimT}
\safemath{\rvecu}{\bimU}
\safemath{\rvecv}{\bimV}
\safemath{\rvecw}{\bimW}
\safemath{\rvecx}{\bimX}
\safemath{\rvecy}{\bimY}
\safemath{\rvecz}{\bimZ}

\safemath{\rvecxi}{\bmxi}
\safemath{\rveclambda}{\bmlambda}
\safemath{\rvecmu}{\bmmu}
\safemath{\rvectheta}{\bmtheta}
\safemath{\rvecphi}{\bmphi}

\safemath{\rmatA}{\bimA}
\safemath{\rmatB}{\bimB}
\safemath{\rmatC}{\bimC}
\safemath{\rmatD}{\bimD}
\safemath{\rmatE}{\bimE}
\safemath{\rmatF}{\bimF}
\safemath{\rmatG}{\bimG}
\safemath{\rmatH}{\bimH}
\safemath{\rmatI}{\bimI}
\safemath{\rmatJ}{\bimJ}
\safemath{\rmatK}{\bimK}
\safemath{\rmatL}{\bimL}
\safemath{\rmatM}{\bimM}
\safemath{\rmatN}{\bimN}
\safemath{\rmatO}{\bimO}
\safemath{\rmatP}{\bimP}
\safemath{\rmatQ}{\bimQ}
\safemath{\rmatR}{\bimR}
\safemath{\rmatS}{\bimS}
\safemath{\rmatT}{\bimT}
\safemath{\rmatU}{\bimU}
\safemath{\rmatV}{\bimV}
\safemath{\rmatW}{\bimW}
\safemath{\rmatX}{\bimX}
\safemath{\rmatY}{\bimY}
\safemath{\rmatZ}{\bimZ}

\safemath{\rmatDelta}{\bimDelta}
\safemath{\rmatLambda}{\bimLambda}
\safemath{\rmatPhi}{\bimPhi}
\safemath{\rmatSigma}{\bimSigma}
\safemath{\rmatOmega}{\bimOmega}
\safemath{\rmatTheta}{\bimTheta}

%% file: macros/standard-macros.tex
\usepackage{amssymb}
\usepackage{amsfonts}
\usepackage{mathrsfs}
\usepackage{xspace}
\usepackage{bm}
\usepackage{fancyref}
\usepackage{textcomp}

\usepackage{multirow}
\usepackage{stmaryrd}

\newenvironment{textbmatrix}{	\setlength{\arraycolsep}{2.5pt}%
								\big[\begin{matrix}}{\end{matrix}\big]%
								\raisebox{0.08ex}{\vphantom{M}}}

\def\be{\begin{equation}}
\def\ee{\end{equation}}
\def\een{\nonumber \end{equation}}
\def\mat{\begin{bmatrix}}
\def\emat{\end{bmatrix}}
\def\btm{\begin{textbmatrix}}
\def\etm{\end{textbmatrix}}

\def\ba#1\ea{\begin{align}#1\end{align}}
\def\bas#1\eas{\begin{align*}#1\end{align*}}
\def\bs#1\es{\begin{split}#1\end{split}} 
\def\bg#1\eg{\begin{gather}#1\end{gather}}
\def\bml#1\eml{\begin{multline}#1\end{multline}}
\def\bi#1\ei{\begin{itemize}#1\end{itemize}}

\newcommand{\tp}[1]{\ensuremath{#1^{T}}} 		
\newcommand{\herm}[1]{\ensuremath{#1^{H}}} 	
\newcommand{\pinv}[1]{\ensuremath{#1^{\dagger}}} 	

\safemath{\dirac}{\delta}					
\safemath{\krond}{\dirac}					

\safemath{\upto}{\uparrow}
\safemath{\downto}{\downarrow}
\safemath{\iu}{j}							
\safemath{\ev}{\lambda}						
\safemath{\hilseqspace}{l^{2}}				
\newcommand{\banachfunspace}[1]{\setL^{#1}}	
\safemath{\hilfunspace}{\banachfunspace{2}}	

\safemath{\SNR}{\textsf{SNR}} 				
\safemath{\PAR}{\textsf{PAR}} 				
\safemath{\No}{N_0}							
\safemath{\Es}{E_s}							
\safemath{\Eb}{E_b}							
\safemath{\EbNo}{\frac{\Eb}{\No}}
\safemath{\EsNo}{\frac{\Es}{\No}}

\DeclareMathOperator{\CHop}{\ensuremath{\opH}} 
\safemath{\tvir}{\rndh_{\CHop}}				
\safemath{\tvtf}{\rndl_{\CHop}}				
\safemath{\spf}{\rnds_{\CHop}}				
\safemath{\bff}{H_{\CHop}}					

\safemath{\ircf}{r_{h}}						
\safemath{\tftvcf}{r_{s}}					
\safemath{\tfcf}{r_{l}}						
\safemath{\bfcf}{r_{H}}						

\safemath{\tcorr}{c_h}						
\safemath{\scf}{c_{s}}						
\safemath{\tfcorr}{c_{l}}					
\safemath{\fcorr}{c_{H}}						

\safemath{\mi}{I}							
\safemath{\capacity}{C}						

\safemath{\normal}{\mathcal{N}}			
\safemath{\jpg}{\mathcal{CN}}			
\safemath{\mchain}{\leftrightarrow}		

\safemath{\dB}{\,\mathrm{dB}}
\safemath{\dBm}{\,\mathrm{dBm}}
\safemath{\Hz}{\,\mathrm{Hz}}
\safemath{\kHz}{\,\mathrm{kHz}}
\safemath{\MHz}{\,\mathrm{MHz}}
\safemath{\GHz}{\,\mathrm{GHz}}
\safemath{\s}{\,\mathrm{s}}
\safemath{\ms}{\,\mathrm{ms}}
\safemath{\mus}{\,\mathrm{\text{\textmu}s}}
\safemath{\ns}{\,\mathrm{ns}}
\safemath{\ps}{\,\mathrm{ps}}
\safemath{\meter}{\,\mathrm{m}}
\safemath{\mm}{\,\mathrm{mm}}
\safemath{\cm}{\,\mathrm{cm}}
\safemath{\m}{\,\mathrm{m}}
\safemath{\W}{\,\mathrm{W}}
\safemath{\mW}{\, \mathrm{mW}}
\safemath{\J}{\,\mathrm{J}}
\safemath{\K}{\,\mathrm{K}}
\safemath{\bit}{\,\mathrm{bit}}
\safemath{\nat}{\,\mathrm{nat}}

\safemath{\define}{\triangleq}			

\safemath{\equivalent}{\sim}
\safemath{\distas}{\sim}					
\safemath{\sdiff}{\Delta}				

\safemath{\reals}{\mathbb{R}}
\safemath{\positivereals}{\reals_{+}}
\safemath{\integers}{\mathbb{Z}}
\safemath{\posint}{\integers_{+}}
\safemath{\naturals}{\mathbb{N}}
\safemath{\posnaturals}{\naturals_{+}}
\safemath{\complexset}{\mathbb{C}}
\safemath{\rationals}{\mathbb{Q}}

\newcommand*{\fancyrefapplabelprefix}{app}		
\newcommand*{\fancyrefthmlabelprefix}{thm}		
\newcommand*{\fancyreflemlabelprefix}{lem}		
\newcommand*{\fancyrefcorlabelprefix}{cor}		
\newcommand*{\fancyrefdeflabelprefix}{def}		
\newcommand*{\fancyrefproplabelprefix}{prop}	
\newcommand*{\fancyrefobslabelprefix}{obs}		
\newcommand*{\fancyrefalglabelprefix}{alg}		
\newcommand*{\fancyrefasmlabelprefix}{asm}	    
\newcommand*{\fancyreftbllabelprefix}{tbl}	    

\frefformat{vario}{\fancyrefseclabelprefix}{Sec.~#1}
\frefformat{vario}{\fancyrefthmlabelprefix}{Thm.~#1}
\frefformat{vario}{\fancyreflemlabelprefix}{Lem.~#1}
\frefformat{vario}{\fancyrefcorlabelprefix}{Cor.~#1}
\frefformat{vario}{\fancyrefdeflabelprefix}{Def.~#1}
\frefformat{vario}{\fancyrefobslabelprefix}{Obs.~#1}
\frefformat{vario}{\fancyrefasmlabelprefix}{Ass.~#1}
\frefformat{vario}{\fancyreffiglabelprefix}{Fig.~#1}
\frefformat{vario}{\fancyrefapplabelprefix}{App.~#1} 
\frefformat{vario}{\fancyrefproplabelprefix}{Prop.~#1}
\frefformat{vario}{\fancyrefalglabelprefix}{Alg.~#1}
\frefformat{vario}{\fancyrefeqlabelprefix}{(#1)}
\frefformat{vario}{\fancyreftbllabelprefix}{Tbl.~#1}

%% file: macros/defs.tex
\safemath{\dictab}{[\,\dicta\,\,\dictb\,]}

\safemath{\ysig}{\bmy}
\safemath{\ysighat}{\hat{\ysig}}
\safemath{\ysigdim}{M}
\safemath{\xsig}{\bmx}
\safemath{\xsigdim}{N}
\safemath{\nx}{n_x}
\safemath{\zsig}{\bmz}
\safemath{\zsigdim}{\ysigdim}
\safemath{\rsig}{\bmr}
\safemath{\Adict}{\bA}
\safemath{\Adicttilde}{\widetilde{\Adict}}
\safemath{\Adictdim}{\outputdim\times\xsigdim}
\safemath{\avec}{\bma}
\safemath{\avectilde}{\tilde{\avec}}
\safemath{\Bdict}{\bB}
\safemath{\Bdicttilde}{\widetilde{\Bdict}}
\safemath{\Cdict}{\bC}
\safemath{\cvec}{\bmc}
\safemath{\Ddict}{\bD}
\safemath{\Ddictdim}{\ysigdim\times\xsigdim}
\safemath{\dvec}{\bmd}
\safemath{\Ddicttilde}{\widetilde{\bD}}
\safemath{\Bonb}{\bB}
\safemath{\bvec}{\bmb}
\safemath{\Bonbdim}{\ysigdim\times\ysigdim}
\safemath{\noise}{\bmn}
\safemath{\noisedim}{\ysigim}
\safemath{\err}{\bme}
\safemath{\errdim}{\ysigdim}
\safemath{\errset}{\setE}
\safemath{\nerr}{n_e}
\safemath{\delop}{\bP_\errset}
\safemath{\delopc}{\bP_{{\errset}^c}}

\safemath{\cplxi}{\imath}
\safemath{\cplxj}{\jmath}

\safemath{\dict}{\matD}
\safemath{\inputdim}{N}		
\safemath{\outputdim}{M}		
\safemath{\sparsity}{S}	
\safemath{\inputdimA}{{N_a}}	
\safemath{\inputdimB}{{N_b}}	
\safemath{\elemA}{{n_a}}	
\safemath{\elemB}{{n_b}}	
\safemath{\resA}{\matR_a}	
\safemath{\resB}{\matR_b}	
\safemath{\subD}{\matS} 
\safemath{\subA}{\matS_a} 
\safemath{\subB}{\matS_b} 
\safemath{\dicta}{\matA} 	
\safemath{\dictb}{\matB} 	
\safemath{\hollowS}{H}
\safemath{\hollowA}{H_a}
\safemath{\hollowB}{H_b}
\safemath{\cross}{Z}
\safemath{\coh}{\mu_d}			
\safemath{\coha}{\mu_a}			
\safemath{\cohb}{\mu_b}			
\safemath{\mubs}{\nu}	
\safemath{\cohm}{\mu_m} 
\safemath{\dictset}{\setD}	
\safemath{\dictsetp}{\dictset(\coh,\coha,\cohb)}	
\safemath{\dictsetgen}{\dictset_\text{gen}}
\safemath{\dictsetgenp}{\dictsetgen(\coh)}
\safemath{\dictsetonb}{\dictset_\text{onb}}
\safemath{\dictsetonbp}{\dictsetonb(\coh)}

\safemath{\leftside}{U}
\safemath{\rightsideA}{R_a}
\safemath{\rightsideB}{R_b}

\safemath{\indexS}{\setI_S} 

\safemath{\nb}{n_b}			
\safemath{\coeffa}{p_i}	
\safemath{\coeffb}{q_j}	
\safemath{\seta}{\setP}		
\safemath{\setb}{\setQ}     
\safemath{\setw}{\setW}	
\safemath{\setz}{\setZ}	
\safemath{\cola}{\veca}		
\safemath{\colb}{\vecb}		
\safemath{\cold}{\vecd}		
\safemath{\inputvec}{\vecx} 	
\safemath{\error}{\vece}	
\safemath{\noiseout}{\vecz} 	
\safemath{\inputvecel}{x}
\safemath{\inputveca}{\vecx_a}
\safemath{\inputvecb}{\vecx_b}
\safemath{\outputvec}{\vecy}	
\safemath{\lambdamin}{\lambda_{\mathrm{min}}}

\safemath{\elltwo}{\ell_2}
\safemath{\ellone}{\ell_1}
\safemath{\ellzero}{\ell_0}
\safemath{\ellinf}{\ell_\infty}
\safemath{\ellinftilde}{\ell_{\widetilde\infty}}
\safemath{\licard}{Z(\coh,\coha,\cohb)}
\safemath{\xsol}{\hat{x}}
\safemath{\xbord}{x_b}		
\safemath{\xstat}{x_s}		
\safemath{\xstatLone}{\tilde{x}_s}
\safemath{\order}{\mathcal{O}} 
\safemath{\scales}{\Theta} 
\safemath{\ones}{\mathbf{1}} 
\safemath{\zeroes}{\mathbf{0}} 
\safemath{\thlone}{\kappa(\coh,\cohb)} 
\safemath{\constoneA}{\delta} 
\safemath{\constoneB}{\epsilon} 
\safemath{\nlarge}{L}				   
\safemath{\sumlarge}{S_\nlarge}
\safemath{\maxlarger}{P_\nlarge}	   
\safemath{\Pzero}{\textrm{P0}}	
\safemath{\Pone}{\textrm{P1}}
\safemath{\vecfir}{\vecw}			 
\safemath{\vecsec}{\vecz}
\safemath{\elvecfir}{w}              
\safemath{\elvecsec}{z}				 
\safemath{\nlargefir}{n}
\safemath{\normout}{\gamma}
\safemath{\auxfun}{h}
\safemath{\supp}{\textrm{supp}}

\safemath{\indexa}{\ell}
\safemath{\indexb}{r}
\safemath{\indexc}{i}
\safemath{\indexd}{j}

\safemath{\project}{P}